\documentclass[
    ,final            
  ]
  {aipproc}

\layoutstyle{6x9}

\begin{document}

\title{ Biased Metropolis Sampling for Rugged Free Energy Landscapes
}

\author{Bernd A. Berg$^{1,2}$}{
  address={$^1\,$Department of Physics, Florida State University,
Tallahassee, FL 32306, USA\\ 
$^2\,$School of Computational Science and Information Technology\\
Florida State University, FL 32306, USA }
}

\begin{abstract}
Metropolis simulations of all-atom models of peptides (i.e. small 
proteins) are considered. Inspired by the funnel picture of Bryngelson 
and Wolyness, a transformation of the updating probabilities of the 
dihedral angles is defined, which uses probability densities from a 
higher temperature to improve the algorithmic performance at a lower 
temperature. The method is suitable for canonical as well as for 
generalized ensemble simulations. A simple approximation to the full 
transformation is tested at room temperature for Met-Enkephalin in 
vacuum. Integrated autocorrelation times are found to be reduced by 
factors close to two and a similar improvement due to generalized 
ensemble methods enters multiplicatively. 
\end{abstract}

\maketitle


\section{Introduction}

Reliable simulations of biomolecules are one of nowadays grand 
challenges of computational science. In particular the problem 
of protein folding thermodynamics starting purely from an aminino-acid 
sequence has received major attention. Until recently the prevailing
view has been that it is elusive to search for the native states with
present day simulational techniques due to limitations of time scale
and force field accuracy. Now a barrier appears to be broken, as 
it was reported~\cite{SNPG02} that large scale distributed computing 
allows to achieve folding of the 23-residue design mini-protein BBA5, 
which is relatively insensitive to inaccuracies of the force field. 
The relaxation dynamics of the computer simulation is found to be in 
good agreement with the experimentally observed folding times and 
equilibrium constants.

Molecular dynamics (MD) technics, for a review see~\cite{FrSm96}, 
tend to be the method of first choice for simulations of biomolecules. 
One of the attractive features of MD is that it allows to follow the 
physical time evolution of the system under investigation.
Nevertheless, there has also been activity based on the Metropolis 
method~\cite{Me53}, which allows only for limited dynamical insights 
and has its strength in the generation of configurations which 
are in thermodynamical equilibrium. A major advantage of Metropolis
simulations of biomolecules is that they allows for updates which 
are large moves when one has to follow dynamical trajectories. Such
updates may help to overcome the kinetic trapping problem, which is due 
to a large number of local minima in the free energy space of typical
biomolecules. Already Metropolis simulations of the canonical 
Gibbs-Boltzmann ensemble may jump certain barriers. Using {\it 
generalized ensembles}, which enlarge the Gibbs-Boltzmann ensemble
and/or replace the canonical weights by other weighting factors,
further progress has been made. For reviews see~\cite{MiSu01,Be02}.

To my knowledge {\it umbrella sampling}~\cite{ToVa77} was the first
generalized ensemble method of the literature. Its potential for 
applications to a wide range of interesting physical problems 
remained for a long time dormant. Apparently, one reason was that 
the computational scientists in various areas shied away from 
performing simulations with an a-priori unknown weighting factor. 
As Li and Scheraga put it~\cite{LiSc88}: ``The difficulty of
finding such a weighting factor has prevented wide applications 
of the ``umbrella sampling'' method to many physical systems.''
This changed with the introduction of the multicanonical 
approach~\cite{BeNe92} to complex 
systems~\cite{BeCe92,HaOk93,HaSc94}. Besides having the luck that
the controversy surrounding its first application got quickly 
resolved by analytical results~\cite{BoJa92}, the multicanonical
approach addressed aggressively the problem of finding reliable 
weights. Nowadays a starter kit of Fortran programs is available 
on the Web~\cite{BeF03}.

The replica exchange or parallel tempering (PT) method~\cite{Ge91,HuNe96}
uses a generalized ensemble which enlarges the canonical configuration
space by allowing for the exchange of temperatures within a set of
canonical ensembles. In this and other generalized ensembles the 
Metropolis dynamics at low temperatures can be accelerated by excursions 
to disordered configurations at higher temperatures. Note that, in
contrast to multicanonical simulations, the probabilities of canonically 
rare configurations are not enhanced by PT. For the purpose of 
simulating biomolecules PT was, e.g., studied in 
Ref.\cite{Ha97,SuOk99,RhPa03}.

In this article we discuss a biased updating scheme~\cite{Be03}, which
enhances already the dynamics of canonical Metropolis simulations and
is easily integrated into generalized ensemble simulations too. The
latter point is illustrated for parallel tempering. The biased updating
scheme is inspired by the funnel picture of protein folding~\cite{BrWo87}. 
At relatively high temperatures probability densities of the dynamical
variables are estimated and used at lower temperatures to enhance
the chance of update proposals to lie in the statistically relevant
regions of the configuration space. In some sense this is nothing
else but an elaboration of the original importance sampling concept 
of Metropolis et al.\cite{Me53}.

This article is organized as follows. All-atom protein models, the 
funnel picture and the rugged Metropolis updating are introduced in the
next section. In the subsequent section
numerical results are presented for the brain peptide Met-Enkephalin.
A short summary and conclusions are given in the final section.

\section{All-Atom Protein Models} \label{protein_models}

Proteins are linear polymers of the 20 naturally occurring amino 
acids. Small proteins are called peptides. The problem of protein 
folding is to predict (at room temperature and in solvents) the 3d 
conformations (native structures) from the sequence of amino acids. 
A conformational energy function models the 
interactions between the atoms (units kcal/mol): 
\begin{equation}
 E_{\rm tot} = E_{\rm es} + E_{vdW} + E_{\rm hb} + E_{\rm tors} 
 + E_{\rm sol}\ .  
\end{equation}
The contributions to the energy function are (charges are in units
of the elementary charge):
\begin{eqnarray} 
{\rm Electrostatic}~~E_{\rm es}  & = & 
\sum_{ij} {332\,q_i\,q_j\over \epsilon\,r_{ij}}\ ,\\
{\rm van\ der\ Waals}~~E_{\rm vdW} & = & \sum_{ij} \left( 
 {A_{ij}\over r_{ij}^{12}} - {B_{ij}\over r_{ij}^6} \right)\ ,\\
{\rm hydrogen\ bond}~~E_{\rm hb} & = & \sum_{ij} \left(
 {C_{ij}\over r_{ij}^{12}} - {D_{ij}\over r_{ij}^{10}} \right)\ ,\\
{\rm torsion}~~E_{\rm tors} & = & \sum_l U_l\, 
\left[ 1 \pm \cos (n_l\,\alpha_l) \right]\ . 
\end{eqnarray}
The solvent accessible surface method allows for an
approximative inclusion of solvent interactions:
\begin{equation} \label{E_sol}
 E_{\rm sol}\ =\ \sum_i \sigma_i\, A_i\ ,
\end{equation}
where $\sigma_i$ is the solvation parameter for atom $i$
and $A_i$  the conformation dependent 
solvent accessible surface area.

\begin{figure}
  \includegraphics[height=.3\textheight]{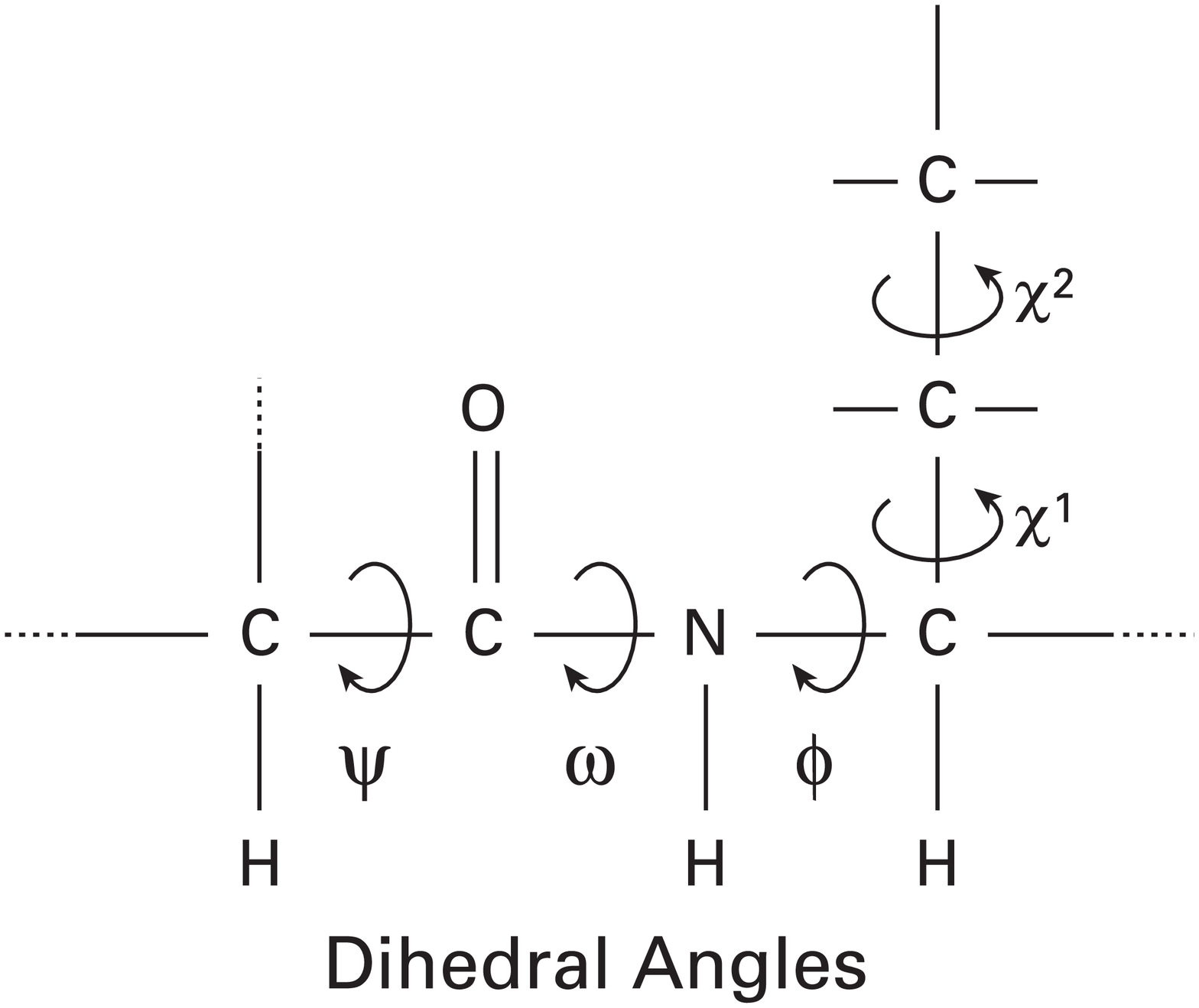}
  \caption{\label{fig_dia} }
\end{figure}

The $r_{ij}$ are the distances between the 
atoms and the $\alpha_l$ are the torsion angles for the 
chemical bonds.  The parameters $q_i,\, A_{ij},\, B_{ij},\, C_{ij},\,
D_{ij},\, U_l$ and $n_l$ are determined from crystal 
structures of amino acids and a number of thus obtained force fields are 
given in the literature.
Bond length and angles fluctuate little and are normally set 
constant.  The important degrees of freedom are the dihedral
angles $\phi, \psi, \omega$ and $\chi$, see Fig.\ref{fig_dia}. 

\begin{figure}
  \includegraphics[height=.3\textheight]{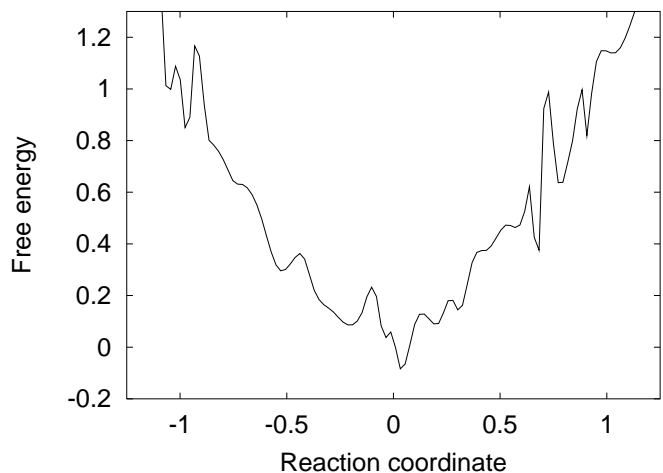}
  \caption{Funnel picture \label{fig_funnel} }
\end{figure}

The funnel picture of Bryngelson and Wolyness~\cite{BrWo87} gives
qualitative insight into the process of protein folding, see
Fig.\ref{fig_funnel}, where a schematic sketch of the free 
energy versus a suitable {\it reaction coordinate} is  given.
However, there is no generical definition of a good reaction 
coordinate, as the funnel lives in the high-dimensional configuration 
space. Here we give a parameter free funnel description~\cite{Be03} 
from higher to lower temperatures, which suggests a method for 
designing the a-priori Metropolis weights for simulations of 
biomolecules.

To be definite, we use the all-atom energy function~\cite{SNS84} 
ECEPP/2 (Empirical Conformational Energy Program for Peptides).
Our dynamical variables $v_i$ are then the dihedral angles, each 
chosen to be in the range $-\pi\le v_i< \pi$, so that the volume 
of the configuration space is $K=(2\pi)^n$.
Let us define the {\it support} of a pd of the dihedral angles.
The support of a pd is the region of configuration 
space where the protein wants to be. Mathematically, we define $K^p$ 
to be the smallest sub-volume of the configuration space for which
\begin{equation} \label{support}
 p = \int_{K^p}\, \prod_{i=1}^n d\,v_i\, \rho (v_1,\dots,v_n;T)
\end{equation}
holds. Here $0<p<1$ is a probability, which ought to be chosen close 
to one, e.g., $p=0.95$. The free energy landscape at temperature 
$T$ is called {\it rugged}, if the support of the pd consists of many 
disconnected parts (this depends of course a bit on the adapted values 
for $p$ and ``many''). 
That a protein folds at room temperature, say $300\,K$, into a
unique native structure $v^0_1,\dots,v^0_n$ means that its pd
$\rho ( v_1,\dots ,v_n; 300\,K)$ describes small fluctuation
around this structure. We are now ready to formulate the funnel 
picture in terms of pds
\begin{equation} \label{rho_r}
\rho_r ( v_1,\dots ,v_n) = \rho ( v_1,\dots ,v_n; T_r),\ 
r=1,\dots,s\ , 
\end{equation}
which are ordered by the temperatures $T_r$, namely
\begin{equation} \label{T_order}
T_1\ >\ T_2\ >\ \dots\ >\ T_f\ .
\end{equation}
The sequence~(\ref{rho_r}) constitutes a protein {\it funnel} when,
for a reasonable choice of the probability $p$ and the 
temperatures~(\ref{T_order}), the following holds:

\begin{enumerate}

\item The pds are rugged.

\item The support of a pd at lower temperature is contained in the 
      support of a pd at higher temperature
\begin{equation} \label{Kp_order}
K^p_1\ \supset\ K^p_2\ \supset\ \dots\ \supset\ K^p_f\ ,
\end{equation}
e.g. for $p=0.95$, $T_1=400\,K$ and $T_f=300\,K$. 

\item With decreasing temperatures $T_r$ the support $K^p_r$ 
      shrinks towards small fluctuations around the native structure.

\end{enumerate}

Properties 2 and 3 are fulfilled for many systems of statistical 
physics, when some groundstate stands in for the native structure.
The remarkable point is that they may still hold for complex 
systems with a rugged free energy landscape, i.e., with property~1 
added. In such systems one finds typically local free energy minima,
which are of negligible statistical importance at low temperatures, 
while populated at higher temperatures. In simulations at low 
temperatures the problem of the canonical ensemble approach is 
that the updating tends to get stuck in those local minima. This 
prevents convergence towards the native structure on realistic 
simulation time scales. On the other hand, the simulations move 
quite freely at higher temperatures, where the native structure 
is of negligible statistical weight. Nevertheless, the support 
of a protein pd may already be severely restricted, as we shall 
illustrate.  The idea is to use a relatively easily calculable 
pd at a higher temperature to improve the performance of the 
simulation at a lower temperature.

The Metropolis importance sampling would be perfected, if we could 
propose new configurations $\{v_i'\}$ with their canonical pd 
$\rho (v'_1,\dots,v'_n;T)$.  Due to the funnel property~2 we expect 
that an {\it estimate} ${\overline \rho} (v_1,\dots,v_n;T')$ from 
some sufficiently close-by higher temperature $T'>T$ will feed useful 
information into the simulation at temperature $T$. The potential for 
computational gains is large because of the funnel property~3. The 
suggested scheme for the Metropolis updating at temperature $T_r$ is 
to propose new configurations $\{v_i'\}$ with the pd
${\overline \rho}_{r-1} (v'_1,\dots,v'_n)$ and to accept them 
with the probability
\begin{equation} \label{P0_acpt}
P_a = \min \left[ 1, \exp \left( - {E'-E\over k\,T_r} \right)\,
{ {\overline \rho}_{r-1} (v_1,\dots,v_n)\over 
{\overline \rho}_{r-1} (v'_1,\dots,v'_n)} \right]\ .
\end{equation}
This equation biases the a-priori probability of each dihedral 
angle with an estimate of its pd from a higher temperature. In
previous literature~\cite{Br85,MHB86} such a biased updating
has been used for the $\phi^4$ theory, where it is efficient to
propose $\phi(i)$ at each lattice size $i$ with its single-site 
probability.

For our temperatures $T_r$ the ordering~(\ref{T_order}) is 
assumed. With the definition 
${\overline \rho}_0 (v_1,\dots,v_n) = (2\pi)^{-n}$
the simulation at the highest temperature, $T_1$, is performed 
with the usual Metropolis algorithm. We have thus a recursive 
scheme, called rugged Metropolis (RM) in the following. When
${\overline \rho}_{r-1} (v_1,\dots,v_n)$ is always a useful 
approximation of $\rho_r (v_1,\dots,v_n)$, the scheme zooms in 
on the native structure, because the pd at $T_f$ governs its 
fluctuations. 

To get things working, we need to construct an estimator 
${\overline \rho} (v_1,\dots,v_n;T_r)$ from the numerical data 
of the RM simulation at temperature $T_r$. Although this is 
neither simple nor straightforward, a variety of approaches offer 
themselves to define and refine the desired estimators. In the 
following we work with the approximation
\begin{equation} \label{rho0_T0}
{\overline \rho} (v_1,\dots,v_n;T_r) = \prod_{i=1}^n
{\overline \rho}^1_i (v_i;T_r) 
\end{equation}
where the ${\overline \rho}^1_i (v_i;T_r)$ are estimators 
of reduced one-variable pds defined by
\begin{equation} \label{pd1}
 \rho^1_i ( v_i; T) = \int_{-\pi}^{+\pi} \prod_{j\ne i} d\,v_j\,
 \rho ( v_1,\dots ,v_n; T)\ .
\end{equation}
The implementation of the resulting algorithm, called RM$_1$, is 
straightforward, as estimators of the one-variable reduced pds are 
easily obtained from the time series of a simulation. The CPU time 
consumption of RM$_1$ is practically identical with the one of the 
conventional Metropolis algorithm.

\section{Numerical Results} \label{results}

To illustrate the developed ideas, we rely on the brain peptide 
Met-Enkephalin, which is a numerically 
well-studied~\cite{LiSc87,OkKi92,HaOk93,MMM94,HaOk03}. 
Met-Enkephalin is determined by the amino-acid sequence 
Tyr--Gly--Gly--Phe--Met or, in the short notation, Y--G--G--F--M, 
where
\begin{eqnarray} \nonumber
{\rm Tyr\ (Y)} &-& {\rm Tyrosine}\\ \nonumber
{\rm Gly\ (G)} &-& {\rm Glycine}\\ \nonumber
{\rm Phe\ (F)} &-& {\rm Phenylalanine}\\ \nonumber
{\rm Met\ (M)} &-& {\rm Methionine}
\end{eqnarray}
Our simulations are performed with a variant of SMMP~\cite{smmp} 
(Simple Molecular Dynamics for Protein) using fully variable $\omega$ 
torsion angles. For the data analysis we keep a times series by 
writing out configurations every 32 sweeps.

\begin{figure}
  \includegraphics[height=.3\textheight]{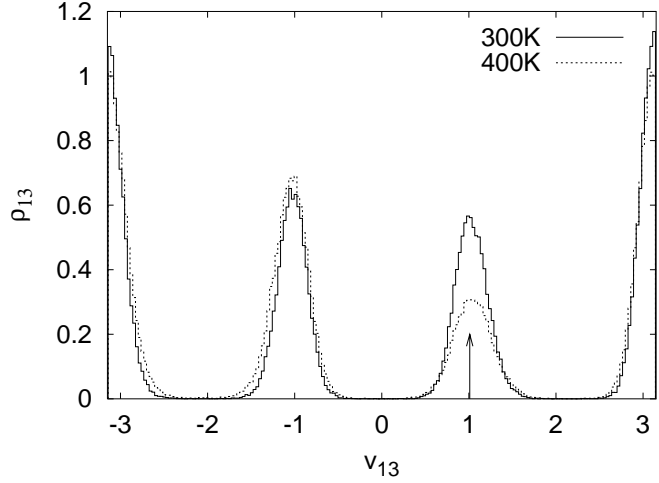}
  \caption{Probability density of a dihedral angle.
  \label{fig_dia13} }
\end{figure}

Fig.\ref{fig_dia13} show the probability densities of the dihedral
angle $v_{13}$ (for the notation see Table~\ref{tab_vac_auto}) at 
400$\,$K and 300$\,$K. This angle is chosen because it illustrates
the possibility of large moves in the Metropolis updating, which
jump barriers of a molecular dynamics simulation. Namely, a 
single update may take us directly from each of the three populated 
regions to each other, whereas one encounters barriers when one
has to move in small increments $\triangle v_{13}$.

To evaluate the relative performance of different algorithms, we
measure the integrated autocorrelation times $\tau_{\rm int}$ for the 
energy and each dihedral angle. The integrated autocorrelation times are 
directly proportional to the computer run times needed to achieve the
same statistical accuracy. For an observable $f$ the autocorrelations 
are 
\begin{equation} \label{Ct}
 C(t) = \langle f_0\,f_t\rangle - \langle f\rangle^2
\end{equation}
where $t$ labels the computer time. Defining $c(t)=C(t)/C(0)$, the
time-dependent integrated autocorrelation time is given by
\begin{equation} \label{tau_int_t}
 \tau_{\rm int}(t) = 1 + 2 \sum_{t'=1}^t c(t')\ .
\end{equation}
Formally the integrated autocorrelation time $\tau_{\rm int}$ is 
defined by $\tau_{\rm int}=\lim_{t\to\infty}\tau_{\rm int}(t)$.
Numerically this limit cannot be reached as the noise 
of the estimator increases faster than the signal. Nevertheless,
one can calculate reliable estimates by reaching a window of $t$ 
values for which $\tau_{\rm int}(t)$ becomes flat, while its error 
bars are still reasonably small. 


\begin{table}[ht]
\caption{ Integrated autocorrelation times.
\protect\label{tab_vac_auto} }
\medskip
\centering
\begin{tabular}{||c|c|c|c|c|c|c|c|c||}                   \hline
$i$&var &res & res      & $400\,K$ & $300\,K$ & $300\,K$
                        & $300\,K$ & $300\,$K \\ \hline
 & & & & Metro & Metro & RM$_1$ & PT & PT$+$RM$_1$\\ \hline
 1.&$\chi^1$&Tyr-1&Tyr-1& 2.16 (08) & 15.8 (2.0) &  9.36 (72)  
                        & 6.28 (46) & 3.38 (22)     \\ \hline
 2.&$\chi^2$&Tyr-1&Tyr-1& 1.23 (02) & 2.96 (25)  &  1.68 (08) 
                        & 1.80 (08) & 1.23 (03)    \\ \hline
 3.&$\chi^6$&Tyr-1&Tyr-1& 1.07 (02) & 2.00 (12)  &  1.58 (09) 
                        & 1.38 (03) & 1.10 (02)    \\ \hline
 4.&$\phi$  &Tyr-1&Tyr-1& 1.49 (05) & 5.77 (48)  &  3.31 (23)  
                        & 1.96 (07) & 1.91 (13)    \\ \hline
 5.&$\psi$  &Tyr-1&Gly-2& 5.02 (14) & 62~~(13)   & 30.3 (2.0)   
                        &15.9 (1.4) & 8.61 (53)     \\ \hline
 6.&$\omega$&Tyr-1&Gly-2& 3.09 (10) & 21.1 (1.8) &  9.68 (66)  
                        & 7.85 (36) & 4.62 (55)    \\ \hline
 7.&$\phi$  &Gly-2&Gly-2& 6.03 (33) & 134 (25)   & 66.6 (7.6)  
                        &26.6 (1.6) & 13.8 (0.6)     \\ \hline
 8.&$\psi$  &Gly-2&Gly-3& 7.49 (50) & 185 (37)   & 91~~(13)
                        &30.6 (2.2) & 18.2 (1.8)     \\ \hline
 9.&$\omega$&Gly-2&Gly-3& 4.50 (15) & 31.4 (2.7) & 14.8 (0.9) 
                        &14.6 (0.7) & 5.51 (31)     \\ \hline
10.&$\phi$  &Gly-3&Gly-3& 7.49 (47) & 167 (27)   & 80.6 (7.0)   
                        &32.7 (3.1) & 22.6 (2.7)     \\ \hline
11.&$\psi$  &Gly-3&Phe-4& 5.05 (30) & 150 (33)   & 81~~(12)  
                        &35.9 (3.5) & 16.7 (0.8)     \\ \hline
12.&$\omega$&Gly-3&Phe-4& 3.33 (10) & 13.53 (90) &  6.71 (56) 
                        & 7.48 (04) & 3.15 (25)    \\ \hline
13.&$\chi^1$&Phe-4&Phe-4& 1.85 (04) & 14.7 (2.7) &  5.51 (70)
                        & 3.29 (16) & 1.71 (06)    \\ \hline
14.&$\chi^2$&Phe-4&Phe-4& 1.18 (03) & 1.77 (08)  &  1.42 (07) 
                        & 1.19 (04) & 1.10 (02)    \\ \hline
15.&$\phi$  &Phe-4&Phe-4& 6.60 (19) & 116 (24)   & 57.9 (4.2) 
                        &30.9 (2.7) & 15.7 (1.2)     \\ \hline
16.&$\psi$  &Phe-4&Met-5& 9.17 (56) & 191 (35)   & 88~~(12)
                        &40.3 (3.2) & 20.8 (1.5)     \\ \hline
17.&$\omega$&Phe-4&Met-5& 1.96 (07) & 7.71 (90)  &  4.57 (41) 
                        & 3.46 (22) & 1.91 (11)    \\ \hline
18.&$\chi^1$&Met-5&Met-5& 1.61 (07) & 10.8 (1.5) &  7.59 (85) 
                        & 4.50 (41) & 3.51 (24)    \\ \hline
19.&$\chi^2$&Met-5&Met-5& 1.19 (02) & 1.68 (05)  &  1.16 (03) 
                        & 1.30 (04) & 1.12 (04)    \\ \hline
20.&$\chi^3$&Met-5&Met-5& 1.01 (01) & 1.05 (02)  &  1.04 (02)  
                        & 1.04 (02) & 1.01 (01)    \\ \hline
21.&$\chi^4$&Met-5&Met-5& 1.00 (01) & 1.01 (01)  &  1.00 (01)  
                        & 1.01 (01) & 1.01 (01)    \\ \hline
22.&$\phi$  &Met-5&Met-5& 2.77 (10) & 31.4 (2.9) & 20.8 (1.9) 
                        &14.9 (1.1) &  9.16 (76)     \\ \hline
23.&$\phi$  &Met-5&Met-5& 1.54 (05) & 21.4 (2.3) & 13.9 (1.7)  
                        & 7.83 (48) & 3.73 (19)     \\ \hline
24.&$\omega$&Met-5&Met-5& 1.06 (01) & 1.14 (02)  &  1.03 (02) 
                        & 1.08 (01) & 1.03 (02)    \\ \hline
   & $E$    &     &     & 4.98 (20) & 49.6 (5.0) & 26.2 (1.6) 
                        &19.9 (1.6) & 9.94 (60)    \\ \hline
\end{tabular} \end{table} \vspace*{0.2cm}

Columns~5 and~6 of Table~\ref{tab_vac_auto} collect our estimates
of the integrated autocorrelation times from conventional,
canonical Metropolis simulations at 400$\,$K and 300$\,$K. We 
find a remarkable slowing down. For the energy, which is 
characteristic for the over-all behavior, $\tau_{\rm int}$ 
increases by a factor of about ten. For certain angles, e.g.
$v_{10}$, $\tau_{\rm int}$ increases even by factors larger than 
twenty. At 400$\,$K conventional Metropolis simulations allow to 
calculate observables with relative ease, while this is no longer 
the case at 300$\,$K.  Our aim is to use information from the 
400$\,$K simulation to improve on the performance at 300$\,$K.

\begin{figure}
  \includegraphics[height=.3\textheight]{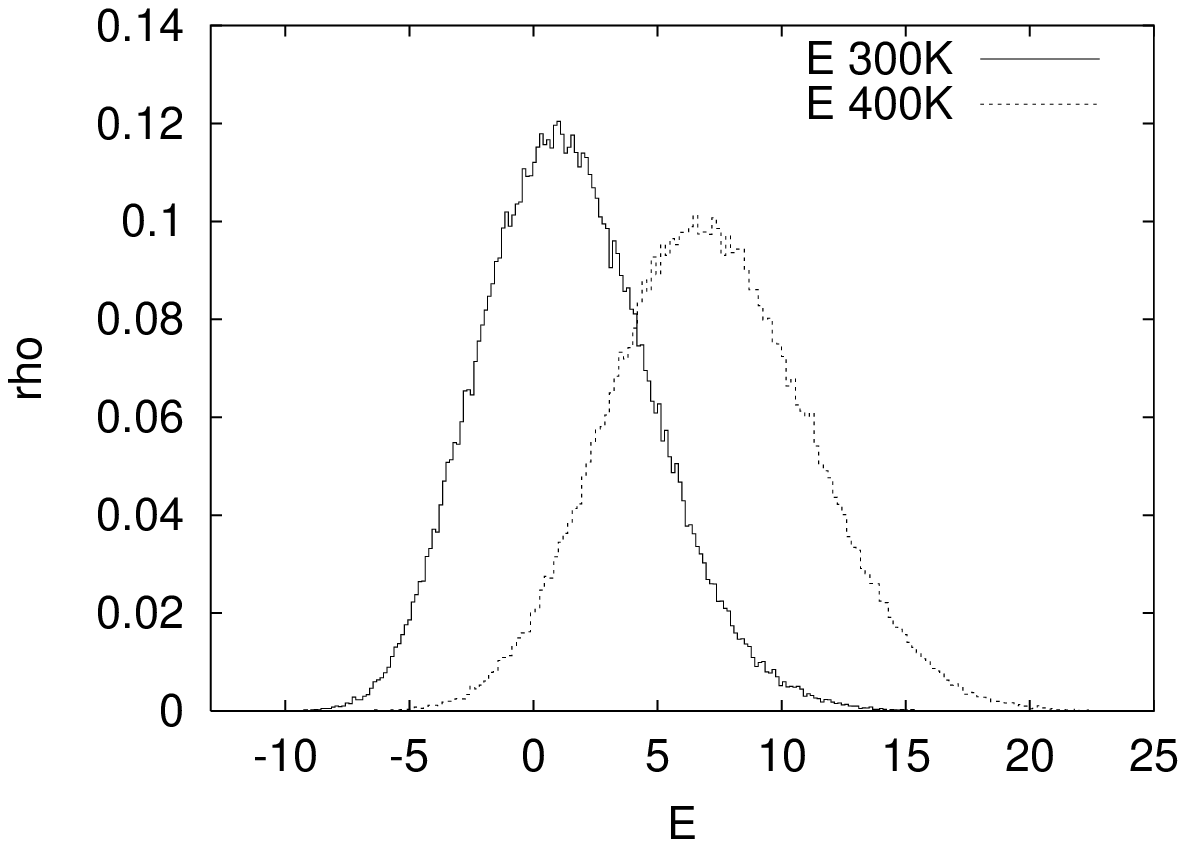}
  \caption{Met-Enkephalin (internal) energy histograms at 300$\,$K 
  and 400$\,$K. \label{fig_energy} }
\end{figure}

One finds that the energy histograms of the 400$\,$K and the 
300$\,$K simulation have a considerable overlap, see 
Fig.\ref{fig_energy}. Therefore, the two temperatures are
well-suited to be combined into a PT~\cite{Ge91,HuNe96} simulation.
We abstain here from introducing additional temperatures, because 
our aim is to get a clear understanding of the improvement of the 
300$\,$K simulation due to PT input from 400$\,$K.

\begin{figure}
  \includegraphics[height=.3\textheight]{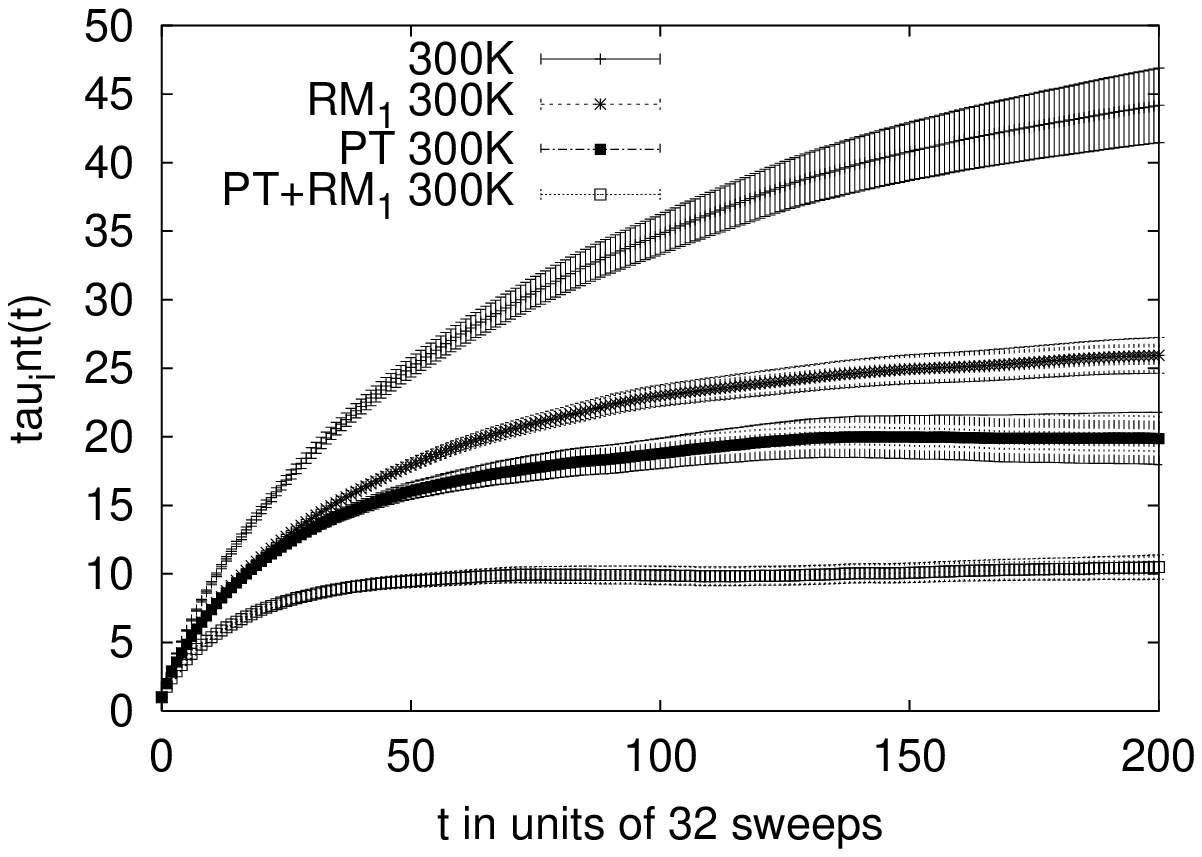}
  \caption{Integrated autocorrelation times for the energy
           at 300$\,$K.  \label{fig_auto} }
\end{figure}

In Fig.\ref{fig_auto} integrated autocorrelations times for the energy
variable are compared at 300$\,$K. The order of the curves agrees with 
the order in the figure legend. From up to down: Conventional
canonical Metropolis simulation, RM$_1$ improved canonical Metropolis
simulation with input from 400$\,$K, PT simulation coupled to 400$\,$K 
and the RM$_1$ improved PT simulation. The variable $t$ of 
equation~(\ref{tau_int_t}) is given in units of 32 sweeps due to the 
way our data are recorded. In the range shown, i.e. up to 
$t=200\times 32$ sweeps, 
a window exist for the PT and the RM$_1$ improved PT
simulations, which allows to estimate the integrated autocorrelation
times of these simulations. For the other two simulations one has
to go to even larger $t$ values, but it remain possible to estimate
$\tau_{\rm int}$. For the energy and all 24 dihedral angles the
$\tau_{\rm int}$ estimates are collected in Table~\ref{tab_vac_auto}.

For the energy we find a decrease from $\tau_{\rm int}\approx 50$
to $\tau_{\rm int}\approx 25$ due to the RM$_1$ improvement of
the canonical Metropolis simulation, i.e. approximately a factor 
of~2. The PT improvement of this simulation is even larger, namely 
from $\tau_{\rm int}\approx 50$ to $\tau_{\rm int}\approx 20$, i.e.
approximately a factor of 2.5. In both cases the CPU time spent at
400$\,$K is not part of the equation. This is justified as one
should anyway understand the system first at temperatures where 
one does not suffer from long autocorrelation times. For PT the 
factor 2.5 is the improvement in real time when two identical PC 
nodes with some parallel software like MPI (Message Passing 
Interface) are available.  Most remarkably, the two improvements 
multiply: For the RM$_1$ improved PT simulation the autocorrelation 
time is down to $\tau_{\rm int}\approx 10$, only about one more
factor of two away from the $\tau_{\rm int}$ value at 400$\,$K.

Inspection of the $\tau_{\rm int}$ values for the 24 dihedral 
angles shows large differences. For the conventional canonical 
simulation the range is 
from\footnote{ A value $\tau_{\rm int}=1$ means that our resolution 
of measurements every 32 sweeps is too crude to show the 
autocorrelations, which one may still expect on the scale 
of a few sweeps.}
$\tau_{\rm int}\approx 1$ for $v_{21}$ to 
$\tau_{\rm int}\approx 200$ for $v_{16}$. For the PT-RM$_1$ 
simulation it becomes reduced to a range from $\tau_{\rm int}\approx 1$
to $\tau_{\rm int}\approx 20$. This suggests that it is not efficient
to simulate by sweeps, where each dihedral angle is updated once
per sweep. Instead, an algorithm (systematic or random) where the
number of updates per angle is proportional to its integrated
autocorrelation time is expected to be more efficient. Obviously,
this can be implemented without major changes of the existing 
code, but tests of this idea have not yet been completed.
After all these improvements the 300$\,$K simulation is expected
to be no more autocorrelated than the conventional, canonical 
simulation at 400$\,$K. 

All-atom Metropolis simulations of small biomolecules deserve a 
place on their own in the arsenal of computational biophysics.
A good understanding of smaller pieces of a larger protein is one 
of the ingredients, which can pave the way towards the understanding
of large systems like proteins. Major algorithmic problems remain 
to be solved before Metropolis simulations of all-atom models may 
become directly suitable for applications like the folding of 
small to medium sized proteins. 

One algorithmic problem is that of correlated moves of two or more 
dihedral angles. Such moves promise to overcome (jump) essential free 
energy barriers in the case of larger molecules. Within conventional 
canonical Metropolis simulations the acceptance rates for simultaneous
moves of two and more angles are prohibitively small. The RM concept
promises major inroads, but details have not yet been tested out.

An even more challenging problem is the inclusion of solvent effects. 
Here the large moves of Metropolis updates create the {\it cavity 
problem}. We need a cavity in the surrounding water to accommodate the 
move and updates of the dihedral angles do not create one. This appears 
to be one of the major reasons why large scale simulations of proteins 
with a surrounding solvent are to an overwhelming extent done within 
the MD framework, where the cavity problem is avoided due to joint,
small moves of all degrees of freedoms. Within the Metropolis approach 
effective solvent models may provide a solution. They are, e.g., defined 
by a solvent contribution like $E_{\rm sol}$ in Eq.(\ref{E_sol}). 
Unfortunately, no reliable determination of the parameters entering 
this equation exists presently in the literature, as was demonstrated 
by recent simulations~\cite{Ha03,BeHs03}. However, it appears to be 
within the reach of Metropolis simulations to lead the way towards 
determining reliable parameterizations.

\section{Summary and Conclusions} \label{summary}

\begin{itemize}

\item The RM$_1$ approximation to the rugged Metropolis (RM)
method~\cite{Be03} leads already to considerable improvements over 
conventional Metropolis simulations of Met-Enkephalin at 300$\,$K. 
As RM$_1$ is easily implemented and needs no additional computer time, 
it should be used whenever Metropolis simulations of suitable systems 
are done.

\item For larger systems one-variable moves alone will not work
due to correlations between the dihedral angles. The RM approach 
promises sufficiently large acceptance rates for multi-variable 
moves.

\item Ultimately, each biomolecule of interest may need its own, 
specifically designed, Metropolis algorithm. This task includes to
determine reliable parameters for a solvent model like the one
of Eq.(\ref{E_sol}). 

\end{itemize}

\begin{theacknowledgments}
I am indebted to Yuko Okamoto for many useful discussion and to
Robert Swendsen for kindly informing me during this meeting about 
a dynamically optimized Monte Carlo method~\cite{Swendsen}, which 
is tailored for the simulation of biomolecules. Further, I would 
like to thank James Gubernatis and the other organizers of the 
Los Alamos Metropolis workshop for their kind hospitality. This 
work was partially supported by the U.S. Department of Energy 
under contract No. DE-FG02-97ER41022. 
\end{theacknowledgments}

\hyphenation{Post-Script Sprin-ger}

\end{document}